\documentclass[twocolumn]{aastex63}
\usepackage{CJK}
\usepackage{txfonts}
\usepackage{graphicx,natbib,bm,url,color,colortbl}

\newcommand{\EQ}{\begin{equation}}
\newcommand{\EN}{\end{equation}}
\newcommand{\EQA}{\begin{eqnarray}}
\newcommand{\ENA}{\end{eqnarray}}

\newcommand{\Eq}[1]{Eq.~(\ref{#1})}

\newcommand{\Fig}[1]{Fig.~\ref{#1}}

\newcommand{\Tab}[1]{Table~\ref{#1}}

{}
{}
{}

{}
{}
{}
{}
{}
{}
{}
{}
{}
{}
{}
{}
{}
{}
{}
{}
{}
{}
{}
{}
{}

{}
{}
{}

{}

{}
{}

\newcommand{\xx}{\bm{x}}

\def\urms{u_{\rm rms}}

\newcommand{\W}{\,{\rm W}}
\newcommand{\V}{\,{\rm V}}

\newcommand{\T}{\,{\rm T}}
\newcommand{\G}{\,{\rm G}}

\newcommand{\K}{\,{\rm K}}
\newcommand{\g}{\,{\rm g}}
\newcommand{\s}{\,{\rm s}}

\newcommand{\m}{\,{\rm m}}

\newcommand{\J}{\,{\rm J}}

\newcommand{\A}{\,{\rm A}}

\graphicspath{{./fig/}{./png/}}

\def\mrms{{\cal M}_{\rm rms}}

\begin{document}
\begin{CJK*}{UTF8}{gbsn}

\title{Dust growth by accretion of molecules in supersonic interstellar turbulence}
\correspondingauthor{Xiang-Yu Li}
\email{xiangyu.li@pnnl.gov}

  \author[0000-0002-5722-0018]{Xiang-Yu Li (李翔宇)}
\affiliation{Nordita, KTH Royal Institute of Technology and Stockholm University,
10691 Stockholm, Sweden}
\affiliation{Pacific Northwest National Laboratory, Richland, WA 99354, USA}

\author{Lars Mattsson}
\affiliation{Nordita, KTH Royal Institute of Technology and Stockholm University,
10691 Stockholm, Sweden}

\date{\today}

\begin{abstract}

We show that the growth rate of dust grains in cold molecular clouds
is enhanced by the high degree of compressibility of a turbulent, dilute gas. 
By means of high resolution ($1024^3$) numerical simulations, we confirm the theory that the spatial mean growth rate is proportional to the gas-density variance. This also results in broadening of the grain-size distribution (GSD) due to turbulence-induced variation of the grain-growth rate. We show, for the first time in a detailed numerical simulation of hydrodynamic turbulence, that the GSD evolves towards a shape which is a reflection of the gas-density distribution, regardless of the initial distribution. That is, in case of isothermal, rotationally forced turbulence, the GSD tends to be a lognormal distribution.
We also show that in hypersonic turbulence, decoupling of gas and dust becomes important and that this leads to an even further accelerated grain growth. 

\end{abstract}

\section{Introduction}
Dust plays an important role in the formation of stars and planets,
but also the evolution of galaxies throughout the whole history of the Universe.
Moreover, \citet{bernstein2002first} estimated that dust re-radiates about $30\%$ of starlight in the Universe. Therefore, dust regulates emission and absorption of stellar light.

Even at redshifts as high as $z\approx 4-8$ galaxies display high dust fractions,
which requires a very fast dust-formation process and a sudden appearance of dust as soon as there is enough metals to form it \citep{Rowlands14,Mattsson15pre,Mattsson16,Watson15}.
Dust grains are formed mainly by condensation of metals into solid-state material, e.g.,\ silicates, graphite and amorphous carbon \citep[see][and references therein]{mathis1990interstellar,Draine03}.
But accretion of volatiles, forming ``ice mantles'', also accounts for some of the dust mass \citep{jones1994grain}.

Type II Supernovae (SNe) and asymptotic giant branch (AGB) stars not only supply the metals in the cosmic matter cycle, but also a significant amount of dust \citep{valiante2009stellar}. However, the destruction of dust in the interstellar medium (ISM) is significant and therefore a replenishment mechanism is needed \citep{Mattsson11b,Valiante11,Rowlands14}. Growth of dust grains by depleting metals (molecules) in the ISM
is the most efficient in cold molecular clouds (MCs), which is indicated by absorption-line observations \citep[see, e.g.,][]{Savage96a,Jenkins09,DeCia16} 

\citet{hirashita2011effects} showed that the evolutionary history of the GSD is important for understanding the dust abundance in galaxies. A key question in this context is what fraction of the dust mass is in large (micron sized) grains. Dust grains grow by accretion of molecules in MCs and by coagulation of interacting grains \citep{li2020coagulation}.
Thus, growth of dust grains in the ISM is an essential process for the initial phase of planet formation, which is not yet fully understood \citep{johansen2007rapid}.

MCs in the ISM are so dilute that the typical size of dust grains is much smaller than the mean free path of the gas molecules.
\citet{baines1965growth} derived an expression for the accretion, concluding that the mass growth rate of dust grains is proportional to,
and primarily regulated by, the surface area of the grains.
This was done under the assumptions that dust grains are spherical,
a fixed proportion of the colliding gas molecules adhered to the grain and that the gas density is homogeneous.
However, they also showed that the relative velocity of grains moving in the gas plays an important role,
which is interesting in a context of high Mach number turbulence.

Interstellar turbulence displays high Reynolds numbers and is highly compressible with typical Mach numbers larger than 10 \citep{nolan2015density}.
This results in vigorous variations of the gas density, which could potentially change the mass growth rate.
\citet{mattsson2020galactic} included gas-density variation in the accretion process and found that the spatial mean
growth velocity of the dust grain radius scales with the square of the Mach number.

Growth and destruction of dust grains shape the
GSD, which in turn determines the extinction curve of galaxies and
is an important parameter in planet and stars formation  \citep{relano2020evolution}.
\citet{Mathis77} obtained a power law distribution of interstellar dust grains with an exponent of about -3.5  by fitting  observed interstellar extinction in the diffuse ISM. This power law distribution has been widely used and has become the canonical GSD, known as the ``MRN'' distribution.
However, the typical shape of the GSD in MCs is not known.

In this study, we will validate the theory of
\citet[][henceforth ``M20'']{mattsson2020galactic}
with state-of-the-art high resolution numerical simulations of hydrodynamic turbulence including
the accretion process of molecules onto dust grains.
We will show below that the M20 theory agrees
perfectly well with the simulation results.
Based on M20 theory, we will demonstrate that
this grain growth facilitated by gas density
variations leads to a lognormal GSD that is
independent of the initial shape of GSD.

\section{Turbulence and particle dynamics}

\subsection{Momentum equation of interstellar turbulence}
\label{sec:flow}
Interstellar turbulence is governed by the Navier-Stokes equation:
\begin{equation}
	{\partial{\bm u}\over\partial t}+\bm{u}\cdot{\bm{\nabla}}\bm{u}={\bm f}
-\rho^{-1}{\bm{\nabla}} p
	+\rho^{-1}\bm{F}_{\rm visc} ,
\label{turb}
\end{equation}
where ${\bm f}$ is a forcing function \citep[see][]{Brandenburg01},
$p$ is the gas pressure, and $\rho$ is the gas density that follows
the continuity equation,
\begin{equation}
	{\partial\rho\over\partial t}+{\bm{\nabla}}\cdot(\rho\bm{u})=0 .
\end{equation}
The viscosity term $\bm{F}_{\rm visc}^{\nu}$ is given by
\begin{equation}
\bm{F}_{\rm visc}^{\nu}=
 \rho \nu\left({\bm\nabla}^2\bm{u}+\frac{1}{3}\nabla\nabla\cdot\bm{u}+2\bm{{\sf S}}\cdot\nabla\ln\rho\right),
\end{equation}
where $\bm{{\sf S}}={1\over 2}\left[{\bm\nabla} \bm{u} +\left({\bm\nabla} \bm{u} \right)^{T}\right]-{1\over 3}\left({\bm\nabla} \cdot \bm{u} \right){\sf I}$
is the rate-of-strain tensor with ${\sf I}$ the unit tensor
and $\nu$ is the kinetic viscosity of the gas.
However, for high Mach-number simulations, an artificial shock viscosity is needed to smooth the shock such that the shocks can be captured without causing a singularity in the velocity field. With this shock viscosity included, $\bm{F}_{\rm visc}$ becomes a superposition of $\bm{F}_{\rm visc}^{\nu}$
and $\bm{F}_{\rm visc}^{\rm shock}$,
\begin{equation}
  \bm{F}_{\rm visc}=
  \bm{F}_{\rm visc}^{\nu}
  +\rho\zeta_{\rm shock}\nabla\nabla\cdot\bm{u}+\left(\nabla\cdot\bm{u}\right)\nabla\left(\rho\zeta_{\rm shock}\right),
\end{equation}
where the shock viscosity $\zeta_{\rm shock}$ is given by 
\begin{equation}
  \zeta_{\rm shock}=c_{\rm shock}\left<\rm{max}[(-{\bm\nabla}\cdot\bm{u})_+]\right>(\min(\delta x,\delta y,\delta z))^2.
\end{equation}
Here $c_{\rm shock}$ is a constant defining the amplitude of the shock viscosity \citep{haugen2004mach}, $\delta x$, $\delta y$, and $\delta z$, are the lengths of the sides of a mesh cell.

The stochastic solenoidal forcing $\bm{f}$ is given by
\begin{equation}
  \bm{f}(\bm{x},t)=\mbox{Re}\{N\bm{f}_{\bm{k}(t)}\exp[i\bm{k}(t)\cdot\bm{x}+i\phi(t)]\},
\end{equation}
where $\bm{k}(t)$ is the wave vector, $\bm{x}$ is position, and $\phi(t)$ ($|\phi|<\pi$)
is a random phase. The normalisation factor is given by
$N=f_0 c_{\rm s}(kc_{\rm s}/\Delta t)^{1/2}$, where $f_0$ is a
dimensionless factor, $k=|\bm{k}|$, and $\Delta t$ is the
integration time step \citep{BD02}.
In the present study, we choose a completely non-helical
forcing, i.e.,
\begin{equation}
  \bm{f}_{\bm{k}}=\left(\bm{k}\times\bm{e}\right)/\sqrt{\bm{k}^2-(\bm{k}\cdot\bm{e})^2},
\end{equation}
where $\bm{e}$ is the unit vector.

The two dimensionless parameters that characterizes compressible turbulence are the Reynolds number Re and the root-mean-square (rms) Mach number $\mathcal{M}_{\rm rms}$.
Re is defined as \footnote{We note that if Re is defined based on
the energy-injection length scale $L_{\rm inj}$ such that
${\rm Re}^L \equiv u_{\rm rms}L_{\rm inj}/\nu$, ${\rm Re}^L=2\pi Re$. 
Therefore, Re in the present study
is equivalent to $2\pi Re $ in other studies.}
\EQ
{\rm Re} \equiv \frac{u_{\rm rms}}{k_f\nu},
\label{eq:Re}
\EN
where $u_{\rm rms}$ is the rms turbulent velocity
and $k_f$ is the forcing wave number.
$\mathcal{M}_{\rm rms}$ is defined as
\EQ
{\cal{M}_{\rm rms}}=\urms/c_{\rm s},
\label{eq:Ma}
\EN
where $c_{\rm s}$ is the sound speed. We assume that the gas is
isothermal, so that $c_{\rm s}^2=\gamma p/\rho =$~constant, where $\gamma=c_{\rm P}/c_{\rm v}=1$ with the specific heats being $c_{\rm P}$ and $c_{\rm V}$ at constant pressure and constant volume, respectively.

\subsection{Particle dynamics}
The momentum equation for dust grains is given by
\begin{equation}
	\frac{d\bm{x}_i}{dt}=\bm{v}_i
\label{dxidt}
\end{equation}
and
\begin{equation}
	\frac{d\bm{v}_i}{dt}=\frac{1}{\tau_i}(\bm{u}
	-\bm{v}_i)\,.
\label{dVidt}
\end{equation}
The stopping time $\tau_i$ due to the kinetic drag is given by
\begin{equation}
\label{stoppingtime}
\tau_i = \sqrt{\pi\over 8}{\rho_{\rm gr}\over\rho}{a\over  c_{\rm s}} \left(1 + {9\pi\over 128}{|\bm{u}-\bm{v}_i|^2\over c_{\rm s}^2 } \right)^{-1/2},
\end{equation}
where $a$ is the radius of dust grains and $\rho_{\rm gr}$
is the material density of dust grains.
\Eq{stoppingtime} is derived under the assumption that
the particle radius is much smaller than the mean-free-path $\lambda$,
which is the case in most astrophysical contexts \citep[large Knudsen number, Kn~$= \lambda/a \gg 1$][]{armitage2010astrophysics,Mattsson19c}.  
In the limit of low relative Mach numbers ($\mathcal{W} = |\bm{u}-\bm{v}_i|/c_{\rm s}\ll1$),
\Eq{stoppingtime} reduces to
\begin{equation}
\tau_i (\mathcal{W}\ll 1) = \sqrt{\pi\over 8}{\rho_{\rm mat}\over\rho}{a\over c_{\rm s}}.
\end{equation}
When $\mathcal{W}\gg 1$, the correction term in the parenthesis of eq. (\ref{stoppingtime})
is dominating and the momentum equation for dust becomes nonlinear \citep[see][]{Schaaf63,Kwok75,Draine79}.

\subsection{Accretion of gas molecules on dust grains}

\subsubsection{Accretion process}
The rate of change of the grain mass due to accretion of molecules is \citep{mattsson2020galactic}
\EQ
\frac{dm_{\rm gr}}{dt}=4\pi a^2SX_i\bar{u}_t\rho(\xx,t),
\label{dm_dt}
\EN
where $X_i\sim10^{-3}$ is the mass fraction of the relevant growth-species
molecules $i$ in the gas, $S$ is the sticking probability for a molecule
hitting the grain and
\EQ
\bar{u}_t=\sqrt{\frac{8k_BT}{m\pi}}
\label{eq:ut}
\EN
is the thermal mean speed of the molecules. Here, $k_B$ is
the Boltzman constant and $m$ is the molar mass of the molecular gas.
Although $\bar{u}_t$ is dependent on the composition of molecular species, we here focus on how supersonic turbulence affects
the accretion process for a given generic composition. That is, we assume a constant $\bar{u}_t$ for convenience.
As \Eq{eq:ut} can be
written as $\bar{u}_t=\sqrt{8/\pi}c_{\rm s}$,  we shall assume $c_{\rm s}$ is constant
as well, i.e., isothermal conditions.
Since $m_{\rm gr}=\frac{4}{3}\pi a^3\rho_{\rm gr}$, \Eq{dm_dt} can be written in terms of $a$ as
\EQ
\frac{da}{dt}=SX_i\bar{u}_t\frac{\rho(\xx,t)}{\rho_{\rm gr}}.
\label{eq:da/dt}
\EN
We define the growth velocity as $\xi=SX_i\bar{u}_t\frac{\left<\rho\right>}{\rho_{\rm gr}}$. Thus, \Eq{eq:da/dt} can be written as
\EQ
\frac{da}{dt}=\xi(t)\frac{\rho(\xx,t)}{\left<\rho\right>},
\label{eq:da/dt1}
\EN
where $\xi(t)\to 0$ as $t\to \infty$ due to depletion of growth-species molecules. In the following we will assume $\xi =$~constant, however. There are two reasons for this assumption: (1) a constant $\xi$ saves a lot of computation time; (2) we will have a better chance of reaching a regime where effects on the growth rate due to decoupling of gas and dust \citep{baines1965growth} can be seen if the growth is not depletion limited. Assuming $\xi =$~constant may not be entirely realistic, but it is reasonable in an early phase of grain growth and could also be justified by the possibility that undepleted gas can be mixed into the modelled region at a rate similar to that of depletion.

\subsubsection[Theoretical models]{Statistical description}
According to \Eq{eq:da/dt}, with $\xi =$~constant, the change of growth rate is determined by $\rho(\xx,t)$ only.
Using a spatial-mean approach, M20 proposed that the mean growth of dust grains due to accretion is given by
\EQ
\frac{{\rm d}\left<a\right>}{{\rm d}t}=\left<\xi\right>\frac{\left<\rho^2\right>}{\left<\rho\right>^2} = \left<\xi\right> (1+\sigma_\rho^2),
\label{eq:da/dt2}
\EN
considering periodic boundary conditions. Furthermore, the standard deviation $\sigma_s$ of the logarithmic density parameter $s = \ln(\rho/\left<\rho\right>)$ is related to the standard deviation $\sigma_{\rho}$ of the linear gas density,
\EQ
\sigma_\rho^2 = \exp(\sigma_s^2) - 1,
\EN
which in turn is empirically related (via simulations) to the mean Mach number ${\cal M}_{\rm rms}$, although the exact relation can be debated \citep[see, e.g.,][]{Passot98,Lemaster08,Price11,Federrath10}. 
The hypersonic turbulence of cold MCs is therefore indicative of very rapid grain growth according to the M20 theory.

Depletion is omitted in the present study for mainly two reasons.
First, M20 does not consider the effect of dust-gas drift as the
grains become large as predicted by \citet{baines1965growth}.
In order to test the validity of neglecting this drift effect, it
would be ideal if the growth is not depletion-limited since the drift
effect can easily be masked by depletion. Second, simulations with depletion
are computationally very demanding. Adding 
depletion in a physically consistent way is straightforward, but
results in a drastic increase of communication between 
processors. The increase of the computing
time is significant and we would not be able to run a whole suite 
of simulations with our present resources.
 
\subsection{Simulation setup}

\subsubsection{Initial parameters}
We have designed our simulations such that they can match observed dust grain properties and interstellar turbulence, save for the very high Re of ISM. For the sake of convenience, we define a unit length $L_0=3.086\times 10^{18}\, \rm{cm}$, unit velocity $v_0=10^5\,\rm{cm\, s^{-1}}$, and unit mass density
$\rho_0=6.771\times10^{-23}\, \rm{g\,cm^{-3}}$. 
This results in an unit time of $t_0=3.086\times 10^{13}\,\rm{s}\approx10^6$ years.
Unless anything else is stated we adopt the Gaussian system of units (a.k.a. cgs units) in this study and introduce simulation units defined within this system.

To reproduce the  properties of a cold MC we assume an initial mean gas density $\left<\rho\right>_{\rm ini} =49.33\rho_0$
and a sound speed is about $c_s=0.3{\bm v_0}$.
Since the gas (molecular hydrogen) is assumed to obey the ideal gas law, the corresponding temperature is about 30 K.
For the dust we employ a mono-dispersed
and a power-law \citep[MRN][]{Mathis77} initial GSD with a mean initial radius of dust grains
$a_{\rm ini}=10^{-23}L_0$.
The power index is -3.5 such that $f(a,0)$ is in
the form of $f(a,0)=a^{-3.5}$ with cutoff $a_{\rm min}=a_{\rm ini}$ and $a_{\rm max}=3a_{\rm ini}$.
The material density of dust grains is $\rho_{\rm gr}= 3.545\times10^{22}\rho_0 = 2.4$~g~cm$^{-3}$.
We adopt the MRN distribution for the following
reasons. First, even though the MRN distribution is derived from
diffuse ISM instead of dense MCs in the cold ISM,
the {\em initial} GSD of an MCs could in fact be
rather similar to that of the diffuse ISM because
dense MCs are formed from diffuse ISM.
Second, the MRN distribution is often used as the "canonical GSD" in ISM and is considered the natural assumption when no actual constraints exist.

\subsubsection{Numerical method}
We use the {\sc Pencil Code} to solve equations described above, which is a finite difference code designed, in particular, to solve compressible turbulence. A third order Runge-Kutta time stepping is adopted. There are 1953120 individual dust grains (Lagrangian inertial particles) being tracked in the simulations and the growth of dust grains is monitored at each time step. Tracking such a large number of dust grain is numerically demanding, but ensures sufficient statistics.

We simulate a wide range of ${\cal M}_{\rm rms}$ (1.38--11) covering transonic to hypersonic turbulence with similar Reynolds numbers. The simulation with ${\cal M}_{\rm rms}\approx 11$ reaches the typical values of ${\cal M}_{\rm rms}$ observed in large-scale interstellar turbulence \citep{Brunt10,elmegreen2004interstellar}. We run for 9 simulation time units in the case of ${\cal M}_{\rm rms}\approx 11$, using 1024 CPUs and a wall-clock time of $24\times 4=96$ hours. There are 61360 time steps integrated with a time step $dt=5.3\times10^{-5}$, which corresponds to about 21 turnover times. Accretion is turned on when turbulence is well-developed. A summary of all simulations is presented in \Tab{tab:nu}.

\section{Results}

\begin{table*}
\caption{${\rm Re}=100$ is kept the same for all the simulations
by changing $f_0$ and $\nu$. $c_{\rm shock}=2$. $L_x=5$ length unit.
The number of particles is $N_{\rm p}=1953120$.
``cgs'' unit is adopted.}
\centering
\setlength{\tabcolsep}{3pt}
\begin{tabular}{ccccccccccc}
  Run &  $\nu$ & $f_0$ & $N_{\rm grid}$ &$\cal{M}_{\rm rms}$ & ${\rm Re}_{\rm mesh}$ & $\rm{Re}$ & $\bar{\epsilon}$& $\eta$ &$\tau_\eta$ & $\tau_L$ \\
\hline
  A & $1.25\times10^{-3}$ & 1.0  & $512^3$  & 1.38 &10 & 96 & 1.54 & 0.006 & 0.03  & 3.38 \\ 
  B & $2.5\times10^{-3}$ & 4.0  & $1024^3$  & 3.26 &6 & 113 & 24.40 & 0.005 & 0.01  & 1.44 \\ 
  C & $5\times10^{-3}$ & 10.0 & $1024^3$ & 5.97&6  & 104 & 154.73 & 0.005& 0.006 & 0.78 \\ 
  D & $10^{-2}$ & 25.0 & $1024^3$ & 11.03&5  & 96 & 971.27 & 0.006& 0.003 & 0.42 \\ 
\hline
\end{tabular}
\label{tab:nu}
\end{table*}

\subsection{Flow properties at different $\mrms$}
\begin{figure*}\begin{center}
\includegraphics[width=\textwidth]{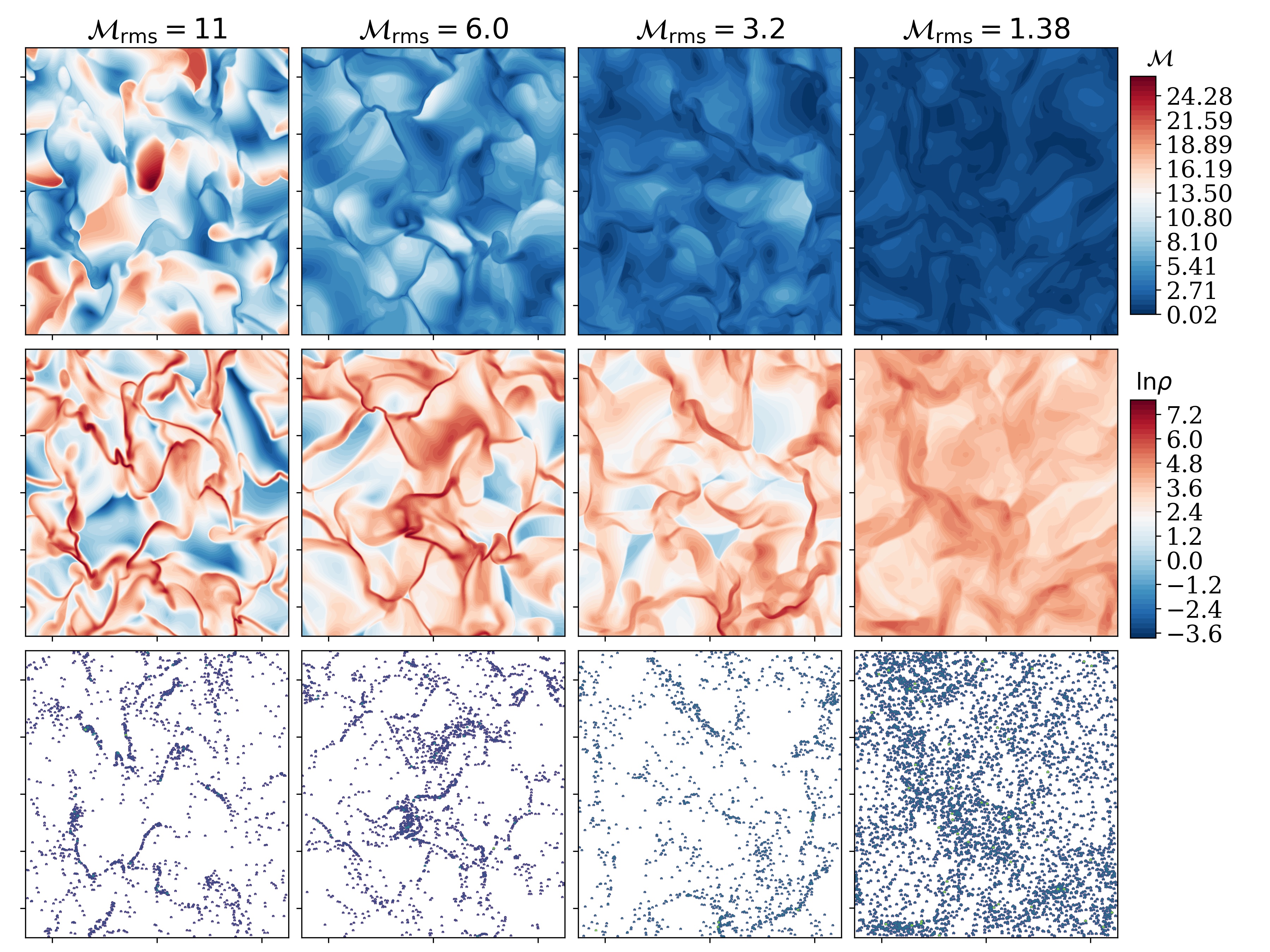}
\end{center}
\caption{Local Mach number ${\cal M} (\xx,t)$,
gas density fields, and number densities of dust grains for simulations with different $\mrms$. Snapshots taken at 8 simulation time units. First row: ${\cal M} (\xx,t)$; second row: logarithmic gas density $\ln[\rho (\xx,t)]$; third row: number densities $n(\xx, t)$ (scatter plots) of dust grains.}
\label{snapshot}
\end{figure*}

\Fig{snapshot} shows snapshots of the local Mach number ${\cal M}(\bm{x},t)$ (upper panel), logarithmic gas density $\ln[\rho(\bm{x},t)]$ (middle panel), and the Eulerian-mapped particle distribution $n(\bm{x},t)$ (lower panel) of four simulations with different ${\cal M}_{\rm rms}$. As ${\cal M}_{\rm rms}$ increases, ${\cal M}(\bm{x},t)$ exhibits more intricate structure and a more inhomogeneous distribution. The same is observed for $\ln[\rho(\bm{x},t)]$, where more pronounced gas filaments are seen at high $\cal{M}_{\rm rms}$. Larger ${\cal M}_{\rm rms}$ also results in stronger spatial clustering of dust grains as shown in the lower panel of \Fig{snapshot}.

\subsection{Comparison with the M20 theory}
\label{sec:M20}

\begin{figure*}\begin{center}
\includegraphics[width=\textwidth]{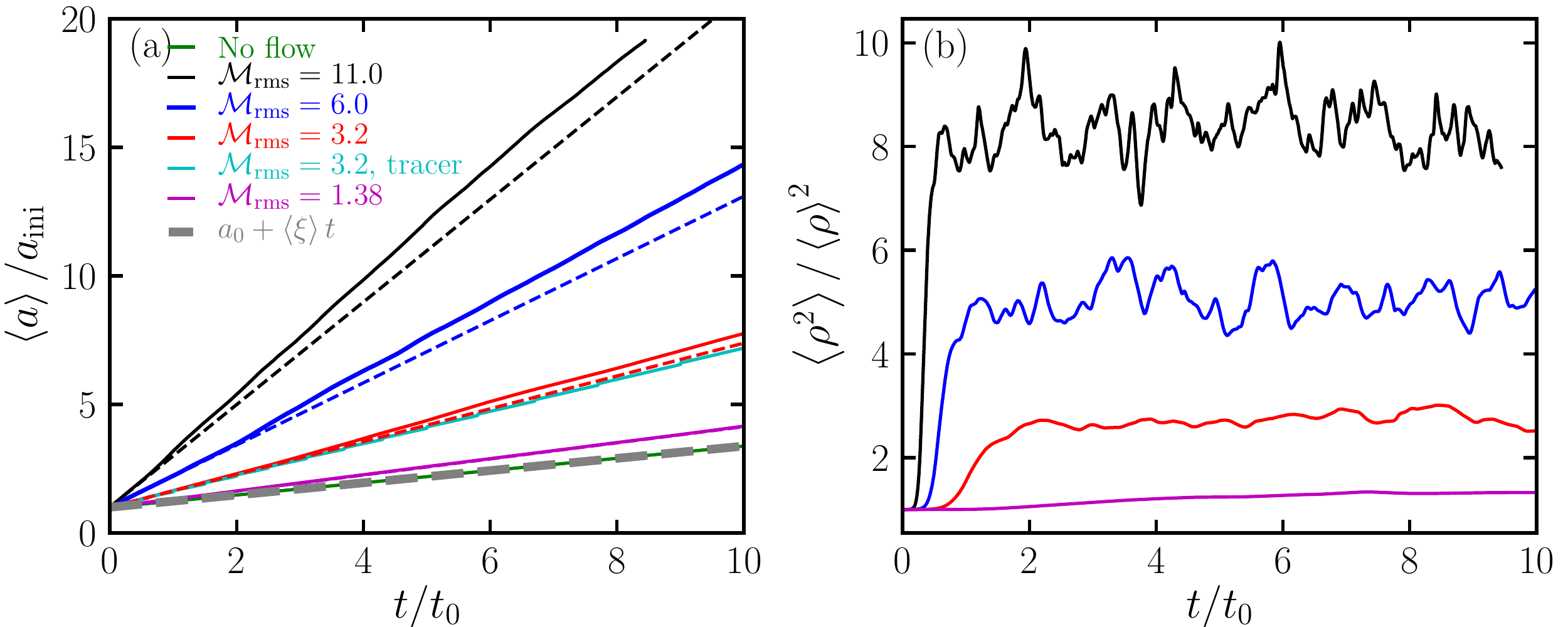}
\end{center}
\caption{Evolution of (a) $\left<a\right>$ and (b)
$\left<\rho^2\right>/\left<\rho\right>^2$ for different ${\cal M}_{\rm rms}$.
Dashed lines in (a) are the corresponding evolution according to \Eq{eq:da/dt2}.
Accretion starts when turbulence is well-developed. The solid cyan curve
represents Run B with dust grains being treated as tracers. The values
of $\overline{\left<\rho^2\right>/\left<\rho\right>^2}$ are 1.31 (magenta curve),
2.71 (red curve), 5.11 (blue curve), and 8.39 (black curve).}
\label{a_comp_massC}
\end{figure*}

M20 assumed that dust grains behave like tracers but can grow due
to accretion as described by \Eq{eq:da/dt2}.
Hence, we started out with a test simulation (corresponding to Run B)
where dust grains are treated as tracers.
\Fig{a_comp_massC}(a) shows that the simulation result (solid cyan line) agrees more
or less perfectly with the theory (red dashed line) in the supersonic regime (${\cal M}_{\rm rms}=3.26$).
The only assumption involved in \Eq{eq:da/dt2} is the periodic boundary conditions, which is required to apply the spatial-mean approach on \Eq{eq:da/dt}. This agreement confirms that the mean growth rate of dust grains is proportional to the square of mean variance of the gas density in the supersonic regime as expressed in \Eq{eq:da/dt2}, provided that dust grains are treated as tracers.

Dust grains are inertial particles and their inertia increase as they grow. Therefore, we have focused on simulations where \Eq{dxidt}--\Eq{stoppingtime} is solved at different ${\cal M}_{\rm rms}$. \Fig{a_comp_massC}(a) shows the evolution of the mean radius $\left<a\right>$ in these simulations. As expected, the spatial mean growth rate of dust grains increases with increasing ${\cal M}_{\rm rms}$. In the transonic regime, the simulation result (solid magenta line) agrees very well with \Eq{eq:da/dt2}. In the supersonic regime and beyond, the simulation results agree with the theory in the beginning, but deviates at later times.
This indicates that \Eq{eq:da/dt2} does not hold in the supersonic regime and beyond,
which was not the case in the tracer-particle limit.
As we can see from \Eq{eq:da/dt2}, the mean growth rate is determined by
the relative variance $\left<\rho^2\right>/\left<\rho\right>^2$,
which is shown in \Fig{a_comp_massC}(b).
Since $\left<\rho^2\right>/\left<\rho\right>^2$ becomes stationary,
the evolution of $\left<a\right>$ as the time integral
of $d\left<a\right>/dt$, is approximately linear with
time as shown in \Fig{a_comp_massC}(a), as predicted by \Eq{eq:da/dt2} if depletion is negligible.

\begin{figure}\begin{center}
\includegraphics[width=0.48\textwidth]{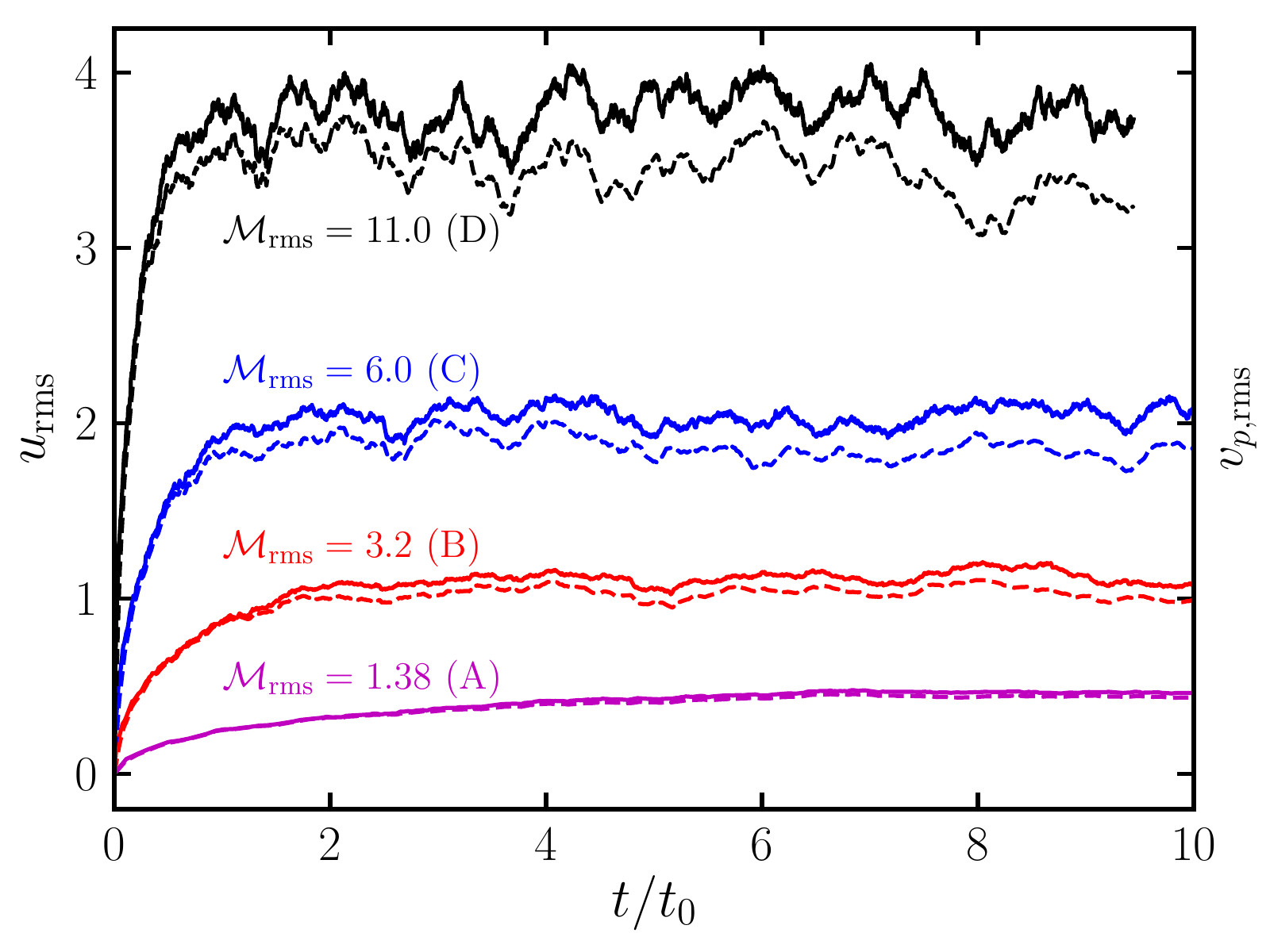}
\end{center}
\caption{The corresponding evolution of $u_{\rm rms}$
(solid lines) and $v_{p,\rm{rms}}$ (dashed lines) for the simulations in \Fig{a_comp_massC}.
}
\label{vp_rms_cali_plot}
\end{figure}

To understand the deviation from the M20 theory, we shall recall that the theory assumed that dust grains are tracers. However, dust grains have inertia, and the kinetics of such particles will
deviate from that of tracers. As shown in \Fig{vp_rms_cali_plot},
$u_{\rm rms}$ and $v_{\rm p,\,rms}$ are very similar in the beginning,
meaning that both gas and dust must be accelerated
to a certain rms velocity before decoupling, at a statistical level,
can actually occur. At later times, $v_{\rm p,\,rms}$ becomes clearly
smaller than $u_{\rm rms}$, indicating that dust grains decouple from
the flow even on average. The decoupling effect is stronger for
large ${\cal M}_{\rm rms}$, which is expected.
Therefore, the deviation from the theory seen
in \Fig{a_comp_massC} is a result of dynamic decoupling of dust grains from the turbulent flow. 

\begin{figure}\begin{center}
\includegraphics[width=0.48\textwidth]{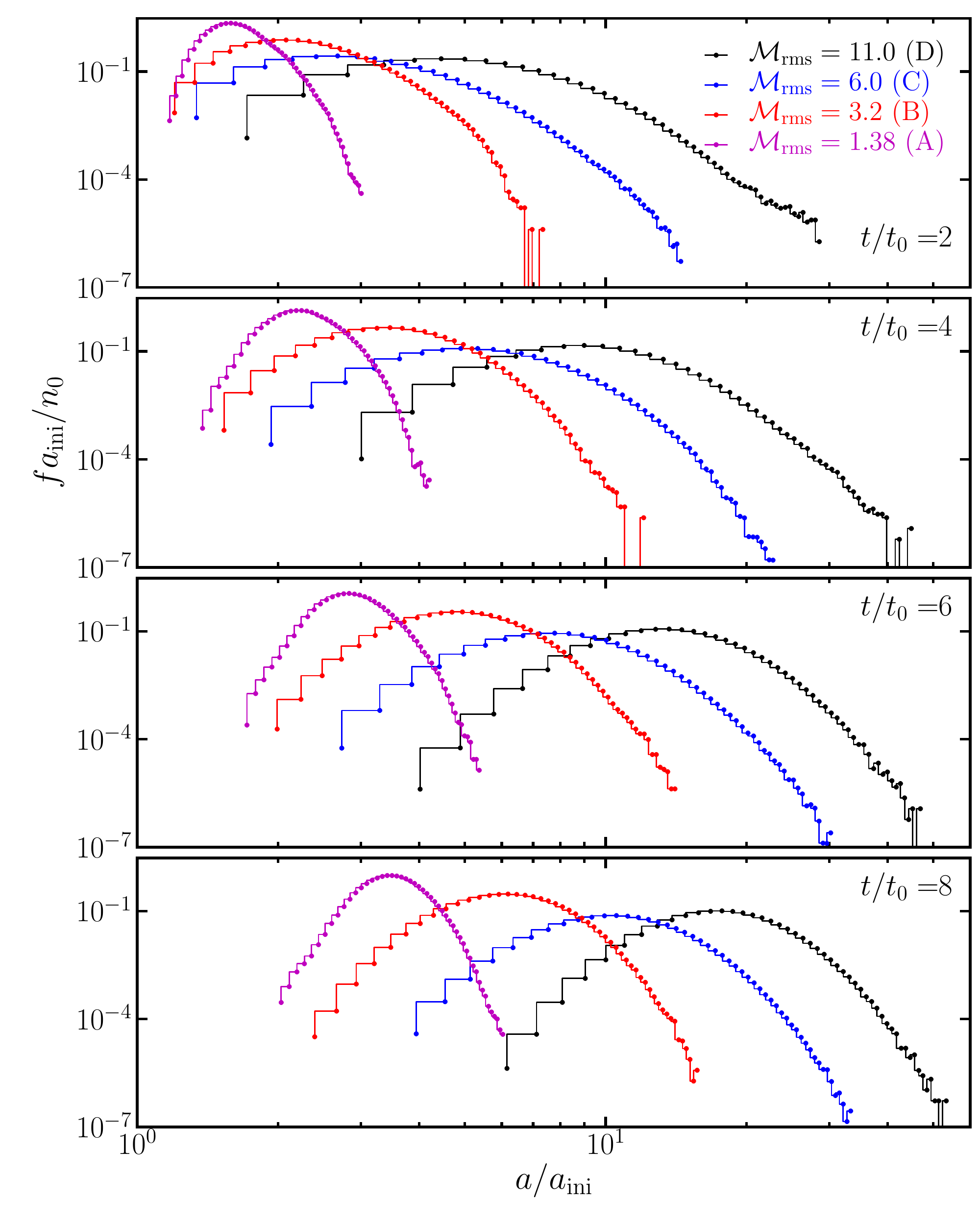}
\end{center}
\caption{The corresponding evolution of $f(a,t)$ for simulations in \Fig{a_comp_massC}.
}
\label{f_evolution_comp_Ma}
\end{figure}

\Fig{f_evolution_comp_Ma} shows GSDs at fixed times (snapshots), where
higher ${\cal M}_{\rm rms}$ results in a broader GSD.
This is because higher ${\cal M}_{\rm rms}$ obviously
leads to stronger variations of $\rho$ as shown in \Fig{snapshot}
and therefore faster growth of dust grains in high-density regions. It is noteworthy that the dust grain population was mono-dispersed initially. 

\begin{figure}\begin{center}
\includegraphics[width=0.48\textwidth]{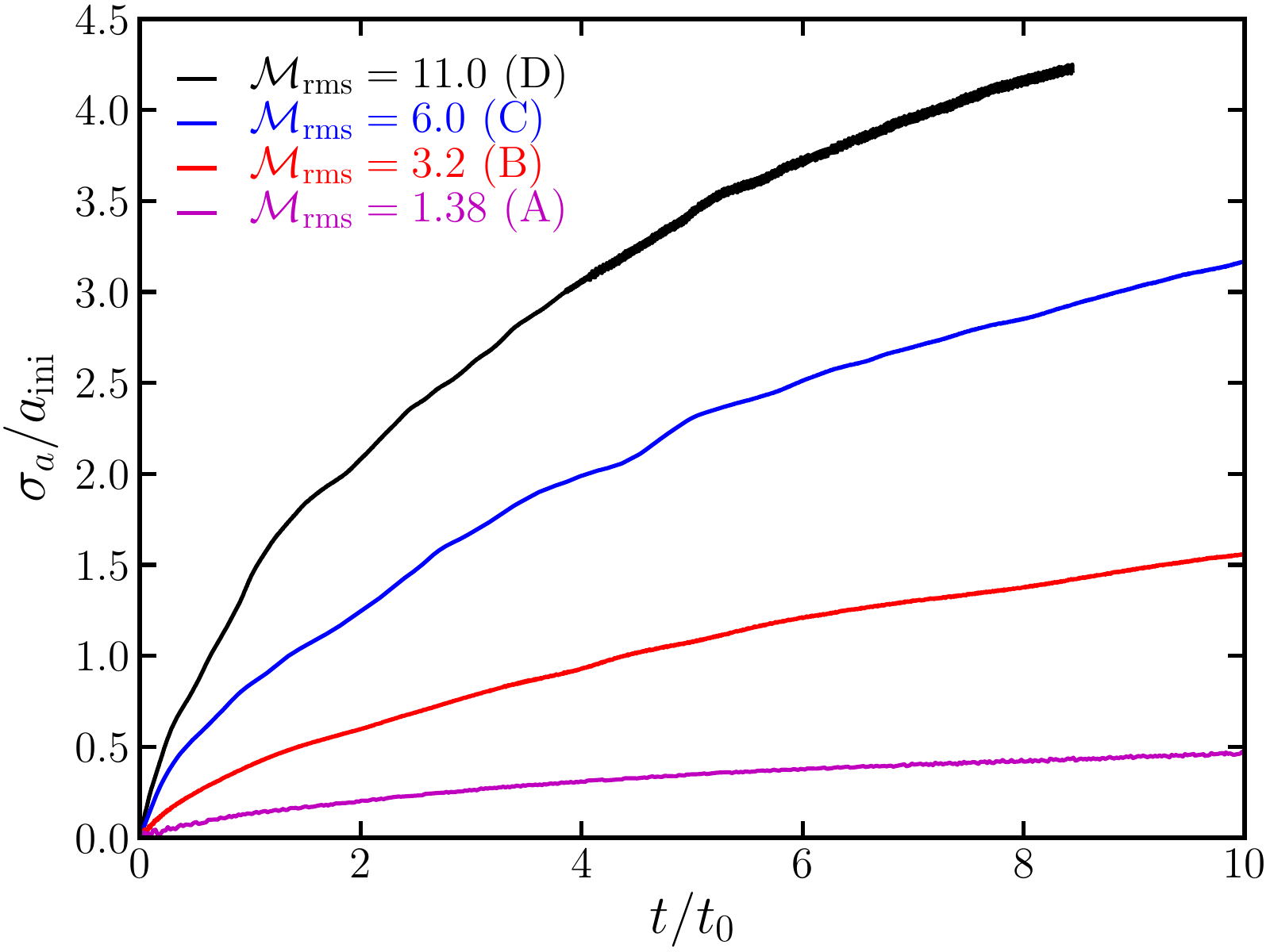}
\end{center}
\caption{The corresponding evolution of $\sigma_a$ for simulations in \Fig{a_comp_massC}.
}
\label{sigma_moments_accretion}
\end{figure}

The evolution of the width/variance of the GSD in terms of $\sigma_a$
is shown in \Fig{sigma_moments_accretion}.
First, we note that $\sigma_a$ grows with time, which is due to
the variations of $\rho$. Second, $\sigma_a$ increases with
increasing ${\cal M}_{\rm rms}$, which demonstrates that
turbulence-induced growth by accretion broadens the GSD.
Note, however, that assuming $\rho =$~constant ($\mrms = 0$, no turbulence),
leads to $\sigma_a = 0$ at all times for an initially mono-dispersed grain population.
Comparing the values of $\sigma_a$ for Run A ($\mrms=1.38$)
and Run D ($\mrms=11.0$) at 8 time units, for example, we find that the
ratio is $\approx 10$. This means that the variance of $f(a,t)$ is enhanced by
a factor of about 10 from transonic flow to the hypersonic flow.

\subsection{Shape of the GSD: a reflection of the gas density PDF?}
\begin{figure}\begin{center}
\includegraphics[width=0.48\textwidth]{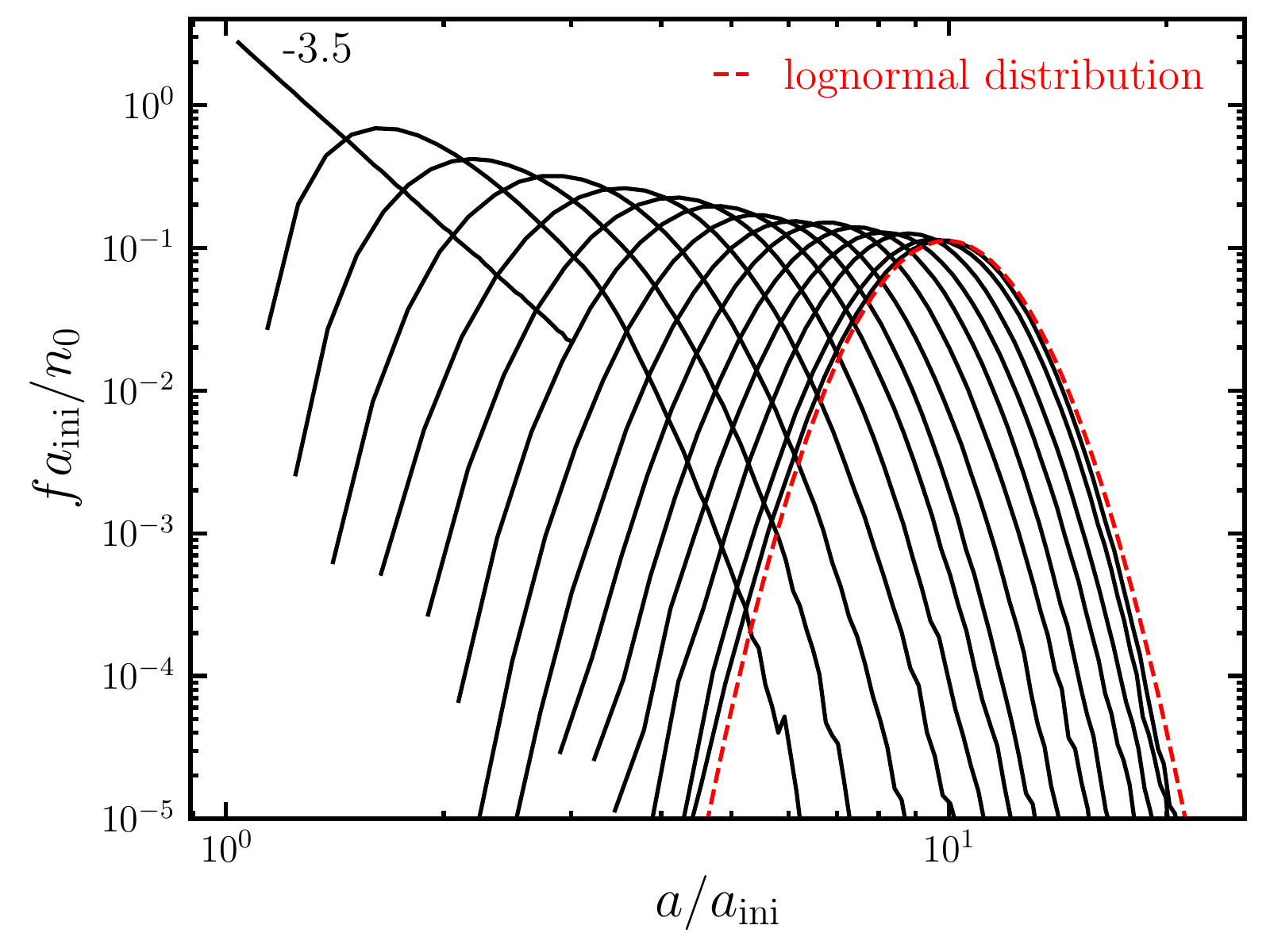}
\end{center}
\caption{Evolution of $f(a,t)$ of Run B
but with a power law initial distribution: $f(a,0)=a^{-3.5}$ with cutoff $a_{\rm min}=a_{\rm ini}$
and $a_{\rm max}=3a_{\rm ini}$. Red dashed line: lognormal distribution with
the same mean and variance. The time interval is every $t_0$.}
\label{f_evolution_SW512_accretion_coag0_nu2p5em3_nus2_f4_L2p5_powerlaw}
\end{figure}

As we mentioned in \ref{sec:M20}, $f(a,t)$ rapidly develops into
a lognormal distribution from a mono-dispersed initial distribution.
\citet{Mathis77} derived a power-law GSD
based on observational constraints in diffuse ISM,
which is now the canonical, widely used GSD known as the ``MRN'' distribution.
\Fig{f_evolution_SW512_accretion_coag0_nu2p5em3_nus2_f4_L2p5_powerlaw} shows the evolution of $f(a,t)$ starting from a power-law GSD with an MRN slope.
A lognormal distribution is the result also in this case, which demonstrates that the lognormal GSD resulting from grain growth in isothermal,
solenoidally forced turbulence is essentially independent of the initial GSD.
To understand the mechanism behind this phenomenon, we shall examine the PDF of the gas density.
The gas-density PDF in our simulations is a lognormal
distribution.
Previous studies \citep[e.g.,][]{Federrath09, Hopkins16, Mattsson18a} of isothermal hydrodynamic turbulence have also generated lognormal distributions of gas density, which can be expressed as
\begin{equation}
\label{eq:rho}
\mathcal{P}(s) = {1\over \sqrt{2\pi}\,\sigma_s}\exp\left[ -{(s - \langle s\rangle)^2\over 2\,\sigma_s^2}\right], \quad s = \ln\left({\rho\over\langle\rho\rangle}\right),
\end{equation}
where $\left<s\right>=\pm\frac{1}{2}\sigma_s^2$ depending on whether the PDF is mass or volume weighted \citep{Li03}. 

The GSD is determined by summing up the local evolution of $da/dt$ in every mesh cell or, which is equivalent, by following the evolution a large number of fluid elements.
The local growth velocity $da/dt$ is proportional to the local density of gas $\rho\left(\xx, t\right)$ as in \Eq{eq:da/dt}. Because $\rho$ follows a lognormal distribution, i.e., \Eq{eq:rho}, it is
very likely that $f(a,t)$ develops into a lognormal distribution. From a more general point of view, it seems that the resultant GSD is a reflection of the gas-density PDF, whatever form this PDF may have.
The mathematical theory behind this is described by Mattsson (2020, submitted).

\section{Discussion}
By applying a spatial mean approach to \Eq{eq:da/dt}, M20 predicted that the mean growth rate of dust grains by accretion is proportional to the mean of the second moment of gas density as described by \Eq{eq:da/dt2}. Simulation results in the present study show that this scaling holds either at early times, when grains are small and their inertia is insignificant, or in the case of small $\mrms$. The analytical theory deviates slightly from the
numerical simulations at high $\mrms$, which we attribute to a higher degree of dynamical decoupling between gas and dust at higher kinetic energies.
We also see that $\left<\rho^2\right>/\left<\rho\right>^2$ becomes stationary and therefore $\left<a\right>$ evolves linearly with time.

Assuming a lognormal gas-density distribution, M20 further proposed that the mean growth rate scales with $\mrms$. However, the semi-empirical nature of the $\sigma_s\,$--$\mrms$ relation combined with the fact that our simulations have a finite and not very high Re, have led us to think it is generally better to compare the M20 theory with our simulations in terms of $\left<\rho^2\right>$ (or $\sigma_\rho$) directly, rather than $\mrms$. The theory predicts an explicit relation with $\sigma_\rho$, while the connection with $\mrms$ is implicit.

We find/confirm that the high degree of compressibility of ISM turbulence can affect grain growth by accretion of molecules in two ways.
On one hand, it increases the mean growth rate of dust grains,
i.e., $d\left<a\right>/dt\propto \left<\rho^2\right>$.
On the other hand, it broadens the GSD as the strong variations
of $\rho(\xx, t)$ leads to a similar variation of the growth rate.
The enhancement is about a factor of 10, going from the transonic
to the hypersonic regime.
This is contrary to the conventional understanding of grain growth,
in which the variance of $f(r,t)$ is unaffected assuming a homogeneous
gas density distribution $\rho(\xx, t)$.
In such a case, the radius $a$ of each grain grows
by the same amount $\Delta a$ in a given time $\Delta t$,
which means the absolute variance of the GSD is conserved.
It is noteworthy that $\sigma_a$ evolves with time and increases with $\mrms$ and $\sigma_\rho$ as a consequence of the accretion process only, in the presence of turbulence. 

Previous work on GSD evolution \citep[e.g.,][]{Hirashita11,Hirashita12,Kuo12,Asano13a,Asano13b,Asano14} have, for the most part, not considered the compressibility and density variations of the ISM, which is particularly strong in MCs. However, more recent work has been directed towards grain processing in simulated gas flows \citep[see, e.g.][]{McKinnon18,Kirchschlager19,Sumpter20}, but we still have a lot to learn. The compressibility of interstellar gas accelerates the mean growth rate and the width of the GSD increases. Therefore, it seems recommendable that gas-density variance due to turbulence should be taken into account in future studies of dust evolution in galaxies, at least in the form of a parametric prescription.

As discussed in \citet{mattsson2020galactic}, the mean growth rate by accretion can be accelerated by an order of magnitude or more if $\mrms\sim 10$, which is quantitatively confirmed by numerical simulations in the present study. Such an enhancement due to the compressibility of interstellar gas makes it possible to deplete essentially all growth-species molecules in an MC onto dust grains within the expected lifetimes of MCs. This indicates that a turbulence-facilitated accretion process can produce a sufficient amount of dust mass to match the observed dust fractions at high redshifts ($z=7-8$), without SNe as extreme dust producers and assuming low dust destruction rates as in, e.g., \citep{Dwek07}. In addition, $\mrms\sim 10$ also leads to roughly a factor of 10 broadening of the GSD. This can potentially accelerate
other dust growth processes such as coagulation \citep{li2020coagulation}, leading to the formation of very large grains ($a> 10\,\mu$m).

Depletion is excluded in this study, but we must emphasize that the depletion time scale is of the same order (or less) as the expected lifetime of a MC.
In the framework of the M20 theory, assuming dust sublimation is negligible, the rate of mass increase of an element $i$ locked up in dust can be expressed
\begin{equation}
    {d\rho_{{\rm d},\,i} \over dt} = {4\pi}\Upsilon_i\,\rho_{{\rm gr},\,i}\,\langle \xi\rangle\, (1+\sigma_{\rm s}^2)\int_0^\infty a^2 f(a)\,da,
\end{equation}
where $\Upsilon_i$ is the mass fraction of $i$ in the generic dust species. The absolute depletion $\delta\rho_i = \rho_i - \rho_{{\rm d},\,i}$ of an element $i$ takes place on timescale which can be estimated as $\tau_{\rm dep} = \delta\rho_i(0)\,|{d\delta \rho_i/dt}|^{-1}$. Assuming a constant $\rho_i$, we also have that $\delta\rho_i(0) = \rho_i$. Thus, $\tau_{\rm dep} \propto (1+\sigma_{\rm s}^2)^{-1}$, which means that the depletion time decreases quickly as the density variance increases. In case of a homogeneous gas medium, the depletion time is to first order just inversely proportional to the gas density, corresponding to  $10^6$--$10^7$ yr for a typical MC \citep[see, e.g.,][]{boland1982carbon}. This is roughly of the same order as the expected lifetime of an MC (see M20). From this we may conclude that many elements can be fully depleted well within the lifetime of a MC and that it may be important to implement depletion in future work.

The GSD is a crucial parameter, as it determines the slope and shape of the ISM extinction curve
and also affects the flux-to-mass ratio and temperature distribution of a dust population \citep[see][and references therein]{Mattsson15}.
We find that the effect of gas-density variance produce a lognormal GSD regardless
of the shape of the initial distribution if the grain-size evolution is dominated by growth from molecule accretion. This is due to the lognormal gas-density distribution generated 
by rotationally forced isothermal turbulence (Mattsson 2020, submitted). 
To our knowledge, there is no directly observational evidence of a lognormal GSD. However, \citet{Weingartner01}
found that the observed galactic emmision is best reproduced
if the size distribution of small dust grains 
is lognormal.

\section{Summary and conclusions}
Numerical simulations covering a wide range of $\mrms$ (1.38--11.0)
have been performed to study the growth of dust grains by accretion of molecules in ISM turbulence. The size evolution of the dust grains was tracked in a Lagrangian manner. Consistent with the theory developed in M20, we have confirmed that the mean growth rate of dust grains is proportional to the second moment of the gas density PDF, i.e., $d\left<a\right>/dt\propto\left<\rho^2\right>$.
In the $\mrms$ range explored here, $\left<\rho^2\right>/\left<\rho\right>^2$
reaches a more or less steady state, which results in a linear mean growth rate of the radius $\left<a\right>$. 

In case of high Mach number turbulence, the simulation results deviate slightly from the theory of \citet{mattsson2020galactic} due to the decoupling of dust grains from the gas,
which, on the other hand, is consistent with the theoretical predictions by \citet{baines1965growth}.
Furthermore, we have found that strong variations of gas density due to turbulence in the ISM leads to significant  broadening of the GSD. This result is indeed noteworthy and, in fact, contrary to the conventional understanding that the absolute width of the GSD is essentially unaffected by growth due to molecule accretion, assuming a homogeneous gas density field. Larger $\mrms$ obviously results in stronger density variations and therefore a wider GSD. 

It is indeed remarkable that the simple M20 theory is able to capture the effects of gas-density variance on the accretion process so accurately. This provides
a possibility to prescribe turbulence accelerated dust growth in models of galactic dust evolution without much loss of generality. Moreover, we found also that when gas density variations were taken into account, the growth process leads to a lognormal GSD, regardless of the shape of the initial distribution. This is a novel result and likely a reflection of the gas-density PDF,
as shown by Mattsson (2020, submitted) in a recent attempt to
construct a mathematical theory of the phenomenon.
As mentioned above, a follow-up study based on this result
in the present study is made by Mattsson (2020, submitted),
which shows that the lognormal
GSD can be derived by modelling turbulent gas-density variation as a Markov process.

The thermal velocity $\bar{u}_t$  of molecular clouds is assumed
to be constant in the present study. However, \Eq{eq:ut} shows that
$\bar{u}_t\propto m^{-1/2}$.
Considering a more realistic chemical composition of MCs
could reduce the growth rate.
But the overall enhancement of the growth rate due to turbulence should not be affected
as it is independent of the chemical composition.
Depletion of growth-species molecules is not included in this study (to save computing time),
but according to \citet{mattsson2020galactic} the predicted net effect is merely to slow down the growth rate.
However, including depletion and other types of grain processing is a currently ongoing project, which we will return to as soon as possible.

\section*{Acknowledgements}
Xiang-Yu Li wishes to thank the Knut and Alice Wallenberg Foundation (grant no. Dnr. KAW 2014.0048) for financial support.
Lars Mattsson wishes to thank the Swedish Research Council
(Vetenskapsrdet, grant no. 2015-04505) for financial support.
Our simulations were performed using resources provided by the Swedish National
Infrastructure for Computing (SNIC) at the Royal Institute of Technology in Stockholm and
Link\"oping University in Linköping.
The Pacific Northwest National Laboratory is operated
for the U.S. Department of Energy by Battelle Memorial Institute under contract DE-AC05-76RL01830.
\software{\sc Pencil Code} is freely
available on \url{https://github.com/pencil-code/}

\end{CJK*}

\end{document}